\documentclass{elsart}
\input epsf.tex
\begin{document}
\runauthor{Adam D. Helfer}
\begin{frontmatter}

\title{State reduction and energy extraction from black holes}

\author{Adam D. Helfer}
\ead{adam@math.missouri.edu}
\address{Department of Mathematics, University of Missouri, Columbia, MO
65211, U.S.A.}

\begin{abstract}
I show that attempts to detect Hawking quanta would reduce  the quantum state
to one containing ultra-energetic incoming particles;  couplings of these to
other systems would extract ultra-high energies from the gravitational
collapse.  As the collapse proceeds, these energies grow exponentially, rapidly
become trans-Planckian, and quantum-gravitational effects must enter.
\end{abstract}
\begin{keyword}
Hawking radiation, black holes, quantum measurement, quantum gravity

\PACS 03.67.Mn, 
03.70.+k, 
04.60.-m, 
04.70.Dy 
\end{keyword}
\end{frontmatter}

Energy, in quantum theory, is measured by the Hamiltonian operator.  Since this
operator generates temporal evolution, energy will be conserved if the operator
remains constant.  But if the quantum system passes through a time-dependent
potential, the Hamiltonian will not be constant, and energy may be exchanged
between the quantum system and the potential.   Two states, whose
energy-contents (precisely, resolutions into energy eigenstates) differ from
each other but little at one time may evolve to have very different
energy-contents at another.

This has consequences for quantum measurements.  The measurement of an operator
may alter the state in a way which seems mild (that is, does not involve very
energetic excitations) at one time, but more substantial at another.  Thus the
measurement --- or the attendant reduction of the state vector --- may induce a
further exchange of energy between the quantum system and the potential.

These apparently reasonable considerations can lead to startling conclusions if
the passage through the time-dependent region induces gross enough changes in
the Hamiltonian.  In this letter, I shall examine what happens in an extreme
case, that of quantum fields passing through a gravitationally collapsing
space--time --- the system famously investigated by Hawking~\cite{Ha74,Ha75}. 
I shall use Hawking's model, but shall be led to substantially different
conclusions from his.

In this case, the time-dependence is the gravitational collapse, and the change
in the Hamiltonian involves both a squeezing transformation (which leads to
particle creation) and a red-shift of field modes passing close to the
incipient black hole.  We shall see that when measurements of Hawking quanta
are made, the state reduces to one in which  quanta which are the {\em
blue-shifted} ingoing precursors of the Hawking quanta are present.  Couplings
of these precursors to other systems will result, when Hawking quanta are
measured, in the extraction of these blue-shifted energies from the collapsing
space--time.

When a black hole forms, these red- and blue-shifts increase exponentially
quickly~\cite{MTW,He01}.    This means that, as the collapse proceeds, the
magnitudes of the energies which might be extracted also grow exponentially. 
We thus have a period in which reduction effects allow larger and larger
energies to be drawn from the collapsing system.

But in short order we face a more fundamental issue.  After a sufficient number
of $\e$-folding times have passed, the magnitudes of the 
energies of the exchanged
quanta have passed the Planck scale.   At this point, attempts to detect
Hawking quanta will inevitably generate Planck-scale, quantum-gravitational,
precursors.  Couplings of the field to other systems mean that these precursors
will mediate Planck-scale, quantum-gravitational, extractions of energy from the
collapsing object.  As we have no theory of quantum gravity, we can say nothing
about what will happen from this point on.

The picture that is drawn here is thus very different from the one drawn by
Hawking.  In Hawking's analysis, the black hole approaches a quasistationary
state in which it can be well-described as a classical object subject to small
quantum corrections resulting in the emission of a weak flux of thermal
radiation.  Here we find that, when the collapse begins, there is a transient
period during which any Hawking quanta which might be produced are entangled
with ultra-high energy excitations of coupled systems, so that the detection of
Hawking quanta would force ultra-high energy exchanges.  However, this period
does not last long. After a number of $\e$-foldings, these ultra-high energies
have passed the Planck scale, and the entire theory (which was predicated on
the neglect of quantum gravity) has broken down.  This result is exciting, for
it means that the black hole becomes an essentially quantum-gravitational
object.  As we shall see, this quantum-gravitational character is not confined
to the event horizon, but influences a neighborhood of the hole, the
probabilities of the effects falling off as a power of the distance.

Related arguments, not involving reduction of the state vector but leading to
the same conclusions, are presented elsewhere~\cite{He04}.

{\em Terminology and conventions.}
In what follows, the distinction between {\em coupling} of quantum systems and
{\em measurement} will be important.  Two systems couple if there are mixing
terms in the total Hamiltonian.  However, we shall reserve the term {\em
measurement} for the action which reduces the quantum state to an eigenstate of
an observable.  We shall use natural units throughout, so that energy,
temperature and frequency are equivalent concepts.

\begin{figure}
\epsfysize 2.5in
\epsfbox{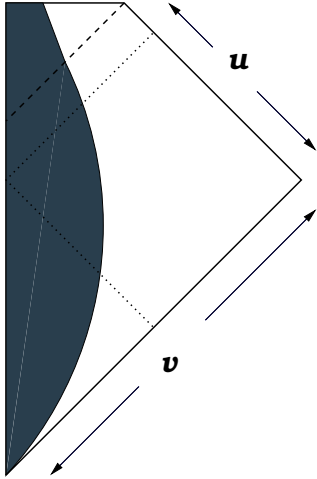}
\caption{A diagram of a black-hole space--time, suppressing angular
variables.  The left-hand edge is the spatial origin. Time increases
upwards, and lines at $45^\circ$ represent the paths of radial light
rays.   The scale has been distorted so that the entire space--time, and some
ideal points at infinity, can be represented.  The region occupied by the
collapsing matter is shaded.  The event horizon is the dashed line; 
the black hole itself is the set of points at and above this. The dotted line
represents a radial light ray beginning at a point on past null infinity, 
moving radially inwards and
passing through the spatial origin (where, in the diagram, it appears to
reflect from the left-hand edge), and escaping to a point on 
future null infinity.}
\end{figure}

Let us recall the basic structure of a space--time modeling the collapse of an
isolated spherically symmetric system of mass $M$ (Fig.~1).   We may introduce
null coordinates $u$ and $v$, the retarded and advanced times (so $u\simeq t-r$
and $v\simeq t+r$ near future and past null infinity, respectively). The
mapping of surfaces of constant phase, that is, the geometric-optics
approximation to propagation, will be important.  For a spherically symmetric
null hypersurface arriving at retarded time $u$ in the future, let $v(u)$ be
the advanced time in the past at which it originated.   That a black hole forms
means that there is a limiting value $v(+\infty )$, which is the advanced time
of formation of the hole. 
(The event horizon is $u=+\infty$.)
Note that if a spherically symmetric wave has period
$\d u$ in the future, its period in the past would be $\d v =v'(u)\, \d u$, so
$v'(u)$ is the red-shift suffered by a spherically symmetric wave passing
through the space--time.  According to gravitational collapse
theory~\cite{MTW,He01}, when a black hole forms one has 
\begin{equation}
v'(u)\simeq \exp -u/(4M)\qquad\hbox{as}\qquad u\to +\infty\, .
\label{eq:asym}
\end{equation}
These exponentially changing red-shifts will play a central role.

Now let $\phi$ be a massless linear Bose field in the space--time.
In the distant past, the field may be resolved into a
sum of normalized field modes 
$p_j$ times annihilation ($a^j$) or creation ($a^*_j$) operators:
\begin{equation}
\phi =\sum _j\left( p_ja^j+{\overline p}^ja^*_j\right) \, .
\end{equation}
Similarly in the distant future, for an appropriate set of field modes
$f_{j'}$ and operators $b^{j'}$, $b^*_{j'}$
\begin{equation}
\phi =\sum _{j'}\left( f_{j'}b^{j'}+{\overline f}^{j'}b^*_{j'}\right)\, .
\end{equation}
The linearity of the field equation implies linear relations between the field
modes (and inversely
the operators):
\begin{equation}
f_{j'}=\sum _j\left(\alpha _{j'}{}^jp_j+\beta _{j'j}{\overline p}^j\right)\, ,
\qquad a^j=\sum _{j'}\left(\alpha _{j'}{}^jb^{j'}+{\overline\beta}^{j'j}
b^*_{j'}\right) \label{eq:Bog}
\end{equation}
where the factors $\alpha _{j'}{}^j$, $\beta _{j'j}$ are known as Bogoliubov
coefficients.  Note that it is the $\beta _{j'j}$ coefficients which mix
creation and annihilation terms; these coefficients are generally responsible
for changes in particle number.

The equations (\ref{eq:Bog})
define an invertible transformation between the field operators
in the distant past and the distant future.  In each of these regimes, there
is a well-defined vacuum (the in-vacuum, characterized by $a^j|0_p\rangle =0$,
and the out-vacuum, by $b^{j'}|0_f\rangle =0$) and well-defined creation and
annihilation operators.  Any physical state may be expressed equally well
as an in-state (given by a sum of $a^*_j$ creation operators acting on
$|0_p\rangle$) or an out-state (given by a sum of $b^*_{j'}$ operators on
$|0_f\rangle$).  We may call these alternative expressions of the state the {\em
past} and {\em future presentations}.

Suppose the field is initially in the in-vacuum state
$|0_p\rangle$.  To understand
what this state would look like if examined at late times, 
that is, its future presentation, we rewrite
the characterizing equation in terms of the operators in the future:
\begin{equation}
a^j|0_p\rangle =\sum _{j'}\left( \alpha _{j'}{}^jb^{j'}
+{\overline\beta}^{jj'} b^*_{j'}\right) |0_p\rangle =0\, .
\end{equation}
It is easily verified that the solution of this is 
\begin{equation}
|0_p\rangle =(\textrm{normalization} ) 
  \exp [(1/2)\sum _{j',k'} Q^{j'k'}b^*_{j'}b^*_{k'}] 
  \, |0_f\rangle\, ,\label{eq:outex}
\end{equation}
where $Q^{j'k'}=-\sum _j (\alpha ^{-1})^{j'}{}_j{\overline\beta} ^{jk'}$.  Here
the various powers of  $Q^{j'k'}b^*_{j'}b^*_{k'}$ create pairs, quadruples, and
so on, of particles.  Thus the past vacuum appears in the future to have a
superposition of various particle numbers:  passage of the field through the
potential has resulted in the creation of particles. Again, note that it is the
possibility $\beta _{j'j}\not= 0$ which allows this production.

In the case of gravitational collapse, the Bogoliubov coefficients were studied
by Hawking.  He found that the primary contributions to $\beta _{jj'}$ came
from specific classes of modes, which we shall denote $j=j_{\rm H}$,
$j'=j'_{\rm H}$, the characteristic ingoing and outgoing Hawking modes.  The
modes $j'_{\rm H}$ have frequencies $\sim T_{\rm H}=(8\pi M)^{-1}$ (the Hawking
temperature), and are maximally compatible with the spherical symmetry.  The
modes $j_{\rm H}$ are the ingoing precursors to $j'_{\rm H}$. They are
concentrated just before the limiting surface $v(+\infty )$,  the advanced time
of formation of the hole.  They are again maximally compatible with the
spherical symmetry, but their frequencies are much higher on account of the
red-shift caused by propagation through the gravitationally-collapsing region:
their frequencies are $\sim T_{\rm H}/v'(u) \simeq T_{\rm H}\exp +u/(4M)$.

The significance of this  exponential growth in the precursors' frequencies has
been a matter of debate~\cite{He03}.  On one hand, the frequencies involved
quickly surpass the Planck scale and thus make the neglect of quantum gravity
problematic.  (This is the {\em trans-Planckian problem}.) On the other, while
ultra-high frequency {\em modes} are indeed implicated in the past, these modes
are unpopulated (the corresponding particle numbers are zero, since the initial
state is vacuum).  Thus many have felt that perhaps with a correct
understanding of the physics one could somehow bypass the awkward use of
ultra-high frequencies. But we shall find that measurement processes can
populate these ultra-energetic modes and transfer exponentially increasing
energies out of the gravitationally collapsing  system.\footnote{A few comments
on one main line of attack on the trans-Planckian problem are in order.  
Jacobson, Unruh {\em et al.}~\cite{Br95,Un95,CJ96,JM00} aim to resolve the
trans-Planckian problem by substituting non-standard rules for propagation of
the fields.   Their goal is to ``insulate'' Hawking's result from
trans-Planckian physics, and this is done by radically altering the actual
mechanism that is used to get that result.  At the moment, work on these
approaches is  ad-hoc and preliminary, and the degree to which they can be said
to eliminate the trans-Planckian problem not entirely clear~\cite{He03}. If,
however, one of these non-standard approaches were to prove correct (that is,
not only self-consistent but the way the world works), then the present
analysis, which relies heavily on the original Hawking model, would have to be
reconsidered.} 

While an exact mathematical treatment in terms of the Bogoliubov coefficients
is possible, in order to present the main ideas without distracting
technicalities we shall just consider the ingoing modes $j_{\rm H}$ and the
outgoing modes $j'_{\rm H}$.  In this approximation, the Bogoliubov
transformation becomes a combination of the effect of the red-shift and a
squeezing transformation.  It red-shifts, and mixes positive and negative
frequencies, but does not convert one frequency into a range of frequencies:
\begin{equation}
a^{j_{\rm H}}\simeq \alpha _{j'_{\rm H}}{}^{j_{\rm H}}b^{j'_{\rm H}}
+{\overline\beta}^{j_{\rm H}j'_{\rm H}} b^*_{j'_{\rm H}}
\, .
\end{equation}
From now on, we shall work only with these modes, and so we shall drop the
postscripts $j=j_{\rm H}$, $j'=j'_{\rm H}$.  Thus the (relevant) Bogoliubov
coefficients will be denoted simply $\alpha$ and $\beta$, and the in-vacuum is
\begin{equation}
|0_p\rangle \simeq (\textrm{normalization} ) \exp [(1/2) Qb^* b^*] \,
  |0_f\rangle\, . \label{eq:outap}
\end{equation}
Hawking found $|\alpha |\sim|\beta |$, so $|Q|=|\overline\beta /\alpha |\sim
1$.

Now let us suppose that the number of particles in the Hawking modes is
actually measured, that is, the operator $n(b)=b^* b$ is measured.  Then the
state reduces to an eigenstate of the operator, with the probability of having
eigenvalue $n$ determined by the state vector, eqn.~(\ref{eq:outap}). (Since
the modes $j'_{\rm H}$ correspond to only a few characteristic periods, and the
characteristic time for emission is typically somewhat greater than
this~\cite{He03},  in fact the most likely value for $n$ is zero.) Of course,
the reduced eigenstate, in the future presentation, is simply $|n_f\rangle
=(n!)^{-1/2} (b^* )^n |0_f\rangle$, the state with $n$ Hawking quanta outgoing.

We may also consider the past presentation of this reduced state.
A routine calculation gives
\begin{equation}
 |n_f\rangle \simeq (\textrm{normalization} ) 
  H_n( (2\overline\alpha \beta )^{-1/2} a^* )\, 
 \exp [(1/2)(\overline\beta /\overline\alpha ) (a^*)^2] \,|0_p\rangle\, ,
   \label{eq:outpastpres}
\end{equation}
where $H_n$ is the $n^{\rm th}$ Hermite polynomial.  Since $a^*$ creates
ultra-energetic particles, we see that the past presentation of the reduced
state is a superposition of ultra-energetic components.  This will be the case
even if no Hawking quanta are detected (that is, $n=0$).  

The upshot of the discussion so far is that the measurement of the
particle-content of field modes of moderate frequencies in the future would
lead to a reduction of the state vector which, in terms of the operators
defined in the past, would contain very energetic particles.  One might think
that, given that the measurement is made in the future, this past presentation
of the state was irrelevant or simply a mathematical way of
speaking without direct physical significance. But this is not so.  
Once the measurement of particle number in the future has been made, the actual
state of the system is the reduced state (until a subsequent measurement is
made).  This means that any device which had in fact been present in the past
and coupled to high-frequency modes would have done so.   

\begin{figure}
\epsfysize 2.5in
\epsfbox{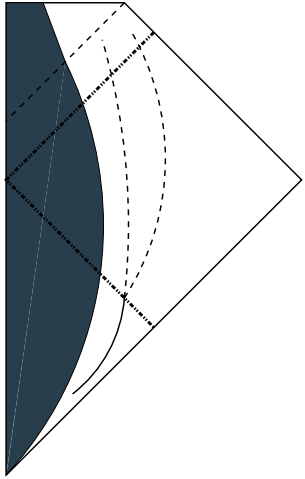}
\caption{Effect of measuring the number of ultra-high frequency quanta absorbed
by a photographic plate, after the number of Hawking quanta has been measured.
The patterned area (shaped $<$) represents the region of space--time
occupied by the measured
Hawking quanta and their precursors.  The forking line
represents two possible trajectories of the plate, according to whether it is
found to have absorbed the 
ultra-energetic precursors (and their energy--momentum) or not.  
(Before measurement of the plate, its state is a superposition of such
trajectories.)  There is a
finite probability that ultra-high energy--momenta will be transferred to the
plate.
}
\end{figure}

To make the discussion more concrete, we will consider a device which we will
refer to as a ``photographic plate'' (sensitive to the ultra-high frequency,
$j_{\rm H}$-mode, quanta) following a trajectory from the distant past to the
distant future, passing through the region potentially occupied by the $j_{\rm
H}$ quanta.  However, whether the device actually is a plate or not (or whether
a real plate could be constructed to respond to the frequencies in question)
will not be important.  What will matter is that the ``plate'' couples to the
operator $n(a)=a^* a$.  

Now suppose we first measure $n(b)$ and then examine the plate.  When we
measure $n(b)$ the state reduces to one with a superposition of ultra-high
frequency quanta.  The plate will couple to those, and so, when it is examined,
it will with positive probability have recorded the passage of the
ultra-energetic $j_{\rm H}$ modes.  And the energy--momentum  of the $j_{\rm
H}$ modes will have  been transmitted to the plate.  (See Fig.~2.)

It should be noted that the plate does not have to be near the collapsing
object for this effect to occur, for the coupling with the $j_{\rm H}$ modes
will occur in a neighborhood of the advanced time of formation of the hole, and
the plate will inevitably pass through this neighborhood.  It is true that the
{\em probability} for any given plate to detect such a mode falls as the radius
at which the plate passes through this neighborhood increases, for the strength
of the mode would fall off like a power of $r$.  However, given that the plate
does detect a quantum, the {\em energy} of that quantum is essentially (apart
from red-shift effects if the plate is very near the collapsing object)
independent of the plate's position.  Thus ultra-high energy--momentum
transfers to the plate are possible, no matter how great its distance from the
hole.

One can make the argument more symmetric by imagining two photographic plates,
one sensitive to the moderate-frequency, $j'_{\rm H}$-mode, Hawking quanta, 
and the other to the ultra-high frequency, $j_{\rm H}$-mode, quanta.  Of
course, these plates are really just ways of visualizing devices coupling to
the number operators $n(b)=b^* b$ and $n(a)=a^* a$, respectively.  We have seen
that if first the Hawking plate is examined, and then the other, the latter
will be found with positive probability to have recorded the passage of
ultra-energetic quanta.  On the other hand, examining the plates in the other
sequence would lead to nothing remarkable.  Examining first the $n(a)$ plate,
one would find no quanta to have been recorded, since the state was the
in-vacuum.  Then examining the Hawking-sensitive plate one might or might not
find Hawking quanta. This order-dependence is a vivid consequence of the fact
that $n(a)$ and  $n(b)$ do not commute.

There is another, important,  perspective.  The coupling of the $n(a)$ plate to
the field results in a coupling of the ultra-energetic $j_{\rm H}$ modes with
the plate.  Since the Hawking ($j'_{\rm H}$) modes are coupled to the $j_{\rm
H}$ modes, this means that the Hawking modes couple with the plate.  However,
since each number state of the Hawking modes corresponds to a superposition of
number states of the $j_{\rm H}$ modes, the number-states of the Hawking modes
are {\em entangled} with states of the plate corresponding to ultra-energetic
disturbances.  Observing a Hawking mode therefore forces the plate into a
superposition of ultra-excited states.

Once this point of view is appreciated, the conventional Hawking out-state
(that is, the future presentation of the state $|0_p\rangle$, eqs.
(\ref{eq:outex},\ref{eq:outap})) appears as a highly non-generic state.  It is
a superposition of different out-particle numbers with coefficients arranged
with a cunning precision in such a way that the couplings to the
ultra-energetic modes in the plate exactly cancel.   These cancellations can be
upset by moderate perturbations of the out-state, as when the number of Hawking
quanta is observed.  We therefore expect that in general couplings of the
quantum field to other systems would result in ultra-energetic entanglements of
those systems.

One natural concern about the thought-experiment presented here is that 
it has the flavor of altering the past (since ultra-energetic
precursors, not present initially, are created).  When we examine this
carefully, however, we shall see that the place for concern seems not
so much with the thought-experiment or its analysis, but with the use
of conventional quantum field theory to describe physics beyond a
certain stage in a gravitationally-collapsing space--time.  And this 
conclusion is precisely the main point of the paper.

Before examining the deeper aspects
of this issue, it is probably well to address a more superficial 
argument against the analysis here which, while motivated by
alteration-of-history concerns, is not really well-founded.
The argument might be put like
this:  ``The initial state is the in-vacuum; this is a basic datum for
the problem; altering this datum corresponds to doing a different
problem.  In other words, the thought-experiment seems to be changing
the rules in the middle of the game.''  This sort of argument
really amounts to denying the possibility of state-vector reduction
as an actual physical process.  It is therefore not really an argument
which is tenable within conventional quantum theory.

Now let us take up the alteration-of-history concerns directly.  We
first point out that there are good general reasons for believing that
the analysis of the thought-experiment presented here will {\em not}
give rise to causal paradoxes.

The past --- as an unalterable historical record --- consists of those classical
data which may be known to an observer together with the results of quantum
measurements already made.  In this paper, what is altered is the quantum state,
and that alteration is by the conventional rules of quantum theory.  So there is
no change in the historical record, and the analysis of the
thought-experiment should be internally consistent against causality
paradoxes, at least to the extent that conventional quantum theory
is.  (This characterization of the historical record will be adequate for
the purposes here, but it becomes
problematic when explicitly quantum-gravitational effects, for
instance superpositions of causal structures, must be considered.)

So there are good general reasons for thinking the analysis of the
thought-experiment is internally consistent.
We shall learn more, however, when we look
more particularly at specific concerns and their resolutions, although
this also requires more extensive discussion.
Let us consider what happens, in the thought-experiment, as the
gravitational collapse proceeds.  The energies which might be
exchanged grow larger and larger.  Is there a point at which these
energies become so large that they affect the known distribution of
energy--momentum, and so the known gravitational field?  Would this
not be an alteration of history?

In answer to this, let us first recall that the neglect of
quantum-gravitational effects must become invalid by the time the
energy-exchanges approach the Planck scale.  Whatever happens beyond
this point is at present a matter of speculation.  For
energy-exchanges below the Planck scale, we must clarify the sense in
which the pre-existing energy--momentum they compete with is
``known.''  Is it a classical known quantity, or one measured
quantum-theoret\-i\-c\-al\-ly?

If, prior to the thought-experiment, the stress--energy is adequately
modeled as a classical quantity, but the reduction effects a
significant enough change that this modeling no longer is valid, then
one does indeed have an alteration-of-history paradox as defined
above.  However, the possible resolutions to this would seem to be
that either there must be new laws of quantum physics forbidding the
sorts of energy-exchanges in the thought experiments, or that the
stress--energy, and hence presumably the space--time geometry, takes
on an essentially quantum character.  Either (or both) of these
resolutions would be in keeping with the main point of this paper,
that conventional quantum field theory must break down in a
gravitationally-collapsing space--time.

If, on the other hand, even prior to the thought-experiment, the
stress--energy must be taken to be a quantum operator, then there is
no alteration-of-history phenomenon as defined above, since one is
simply measuring a sequence of quantum operators (the stress--energy,
and the number operators $n(a)$ and $n(b)$).  Again, since
the stress--energy is the source for Einstein's equation, a quantum
stress--energy implies a quantum gravitational field and a deeper,
quantum-gravitational,
treatment of the entire question is really required.  Again, this is
in line with the main argument here, that quantum gravity must be
considered in the quantum physics of gravitational collapse.

Some final comments are in order.

First, although the discussion has been phrased in terms of ``photographic 
plates,'' there is nothing very special about these.  It was not even important
that these responded precisely to the number operator $n(a)=a^* a$.  What was
really relevant was that the devices coupled to the ultra-energetic field modes
$j_{\rm H}$.  Any such device would be affected by a measurement of the number
of Hawking quanta, because such a measurement will reduce the state to the form
(\ref{eq:outpastpres}), a distribution of the ultra-energetic quanta. If any
coupling to these modes is present, it will be implicated when attempts to
measure the number of Hawking quanta are made.\footnote{In a practical sense it
would be important to know  what sorts of devices can couple to the
ultra-energetic field modes, and with what efficiencies.  Even before the
Planck regime, however, the physics in question becomes speculative, as we do
not really know the behaviors of quantum field theories at extreme energies. 
Here, we are mainly interested in questions of principle, so any coupling,
leading to a positive probability of ultra-energetic effects, is significant.}

Second, I have used the term ``ultra-energetic'' to characterize the implicated
ingoing modes and energy-transfers.  However, this term must be considered
bland in face of the exponentially fast growth of these energies (eq.
(\ref{eq:asym})).  The $\e$-folding time is typically short.  (It is $4M$, the
light-crossing time of the hole, $\simeq 2.0\times 10^{-5}$ s for a solar-mass
hole.)  This means that the energies of the precursors, and the
energy-transfers to the plates, rapidly pass the Planck scale.  At that point,
the theory --- quantum field theory in curved space--time, neglecting quantum
gravity --- has entirely broken down.

This is the main lesson.  When couplings between the quantum field and other
systems are allowed, observation of Hawking quanta will, before long in the
gravitational collapse process, have resulted in Planck-scale energy-transfers 
and therefore it will be impossible to analyze the physics without taking
quantum gravity into account.

The results here imply a more severe breakdown of the  conventional theory of
black-hole radiation than had been contemplated.   Previously, there had been
some concerns about the theory's reliance on trans-Planckian modes, but these
seemed to be only concerns about virtual, vacuum-fluctuation, processes. And
while the neglect of possible quantum-gravitational corrections was considered 
(e.g. refs.~\cite{BM,Ash98}), the primary speculations were that manifestly
quantum-gravitational, Planck-scale, effects would be confined to the hole
itself and leave only secondary imprints, at conventional physical scales,  on
the Hawking process at macroscopic distances. We have seen here, though, that
real Planck-scale effects  at significant distances are possible, and thus the
trans-Planckian problem and manifest quantum-gravitational consequences must be
faced directly.

\end{document}